\documentclass[aps,prl,showpacs,twocolumn]{revtex4-1}
\usepackage{epsfig}
\usepackage{color}

\definecolor{red}{rgb}{1,0,0}
\definecolor{blue}{rgb}{0,0,1}
\definecolor{black}{rgb}{0,0,0}

\newcommand{\blue}{\color{black}}

\newcommand{\be}{\begin{equation}}
\newcommand{\ee}{\end{equation}}
\newcommand{\ba}{\begin{eqnarray}}
\newcommand{\ea}{\end{eqnarray}}

\usepackage{amsmath}
\newlength{\arrow}
\begin{document}
\title{Discontinuous shear thickening without inertia in dense non-Brownian suspensions}

\author{M. Wyart$^1$ and M. E. Cates$^2$  }

\affiliation{$^1$New York University, Center for Soft Matter Research, 4 Washington Place, New York, NY, 10003, USA }
\affiliation{$^2$SUPA, School of Physics and Astronomy, University of Edinburgh, JCMB Kings Buildings, Mayfield Road, Edinburgh EH9 3JZ, United Kingdom }
\date{\today}

\begin{abstract}
A consensus is emerging that discontinuous shear thickening (DST) in dense suspensions marks a transition from a flow state where particles remain well separated by lubrication layers, to one dominated by frictional contacts. We show here that reasonable assumptions about contact proliferation predict two distinct types of DST in the absence of inertia. The first occurs at densities above the jamming point of frictional particles; here the thickened state is completely jammed and (unless particles deform) cannot flow without inhomogeneity or fracture. The second regime shows strain-rate hysteresis and arises at somewhat lower densities where the thickened phase flows smoothly. DST is predicted to arise when finite-range repulsions defer contact formation until a characteristic stress level is exceeded.
\end{abstract}


\maketitle

Shear-thickening, the increase of viscosity $\eta$ with strain rate $\dot\gamma$, is found in many dense suspensions \cite{brown3,barnes}. It has dramatic effects: a person can run across a tank of material that is completely fluid if stirred slowly \cite{brown3}. The details of shear-thickening depend on various factors including Brownian motion, gravity, and inertia; in what follows we consider only large particles in a density-matched viscous fluid, where all three are negligible. Some facts are generic however: (i) If the particle volume fraction $\phi$ is low enough, $\eta$ increases smoothly with strain rate, giving continuous shear thickening (CST). (ii) For large enough $\phi$, a jump of shear stress $\sigma = \eta\dot\gamma$ is instead observed on increasing $\dot\gamma$, giving discontinuous shear thickening (DST). (iii) DST is reversible if $\dot\gamma$ is reduced, but flow-curve hysteresis is apparent \cite{wagner}. (iv) Particle deformability and interactions are important: emulsions and foams do not normally display shear-thickening, nor do  
{\blue strongly attractive particles}
\cite{brown3,barnes}. Short-range interactions can alter the onset density for shear thickening \cite{fernandez}. (v) For fixed particle type, the onset stress (but not strain rate) varies only weakly with $\phi$ \cite{fall,brown3}.

Longstanding explanations for CST and/or DST include the flow-induced formation of hydrodynamic clusters \cite{Cheng,farr,brady2} and shear-induced melting of a partially ordered state \cite{Hoffman}. 
For DST, these
can be set aside for reasons reviewed in \cite{brown3}.
A more promising avenue \cite{barnes}, recently revived \cite{brown1,brown2,brown3}, is that shear-thickening is related to well-known tendency of granular materials to expand under flow (dilatancy). 
In \cite{brown2,brown3} it was proposed that this causes DST under very broad conditions. 

This argument appears too general since it predicts DST for purely hard  particles without inertia. Such particles do show dilatancy: $\phi$ decreases with $\dot\gamma$ at fixed particle pressure $P$ \cite{boyer}.  However dimensional analysis implies that steady state flows in such a system depend on a single parameter  \cite{lemaitreroux}, the viscous number $I_v= \eta_0\dot\gamma/P$, which fixes both $\phi$ and $\sigma/P$. (Here $\eta_0$ is the solvent viscosity.) 
Hence 
at fixed $\phi$ the stress remains linear in strain rate, $\sigma = \eta(\phi)\dot\gamma$, albeit with a viscosity $\eta(\phi)$ that diverges at a jamming density $\phi_m \simeq 0.58$ \cite{boyer}. Beyond $\phi_m$ the system is completely jammed, so homogeneous flow is impossible; here one expects either fracture \cite{cateshaw} or shear-banding with particle migration \cite{besseling}. 

Thus dilatancy at fixed $P$ does not guarantee shear thickening at fixed $\phi$. Instead it can be argued \cite{brown3} that shear thickening arises when the stress exceeds some scale, set by finite interparticle repulsions, at which lubrication films convert to frictional contacts. Such repulsions, by preventing breakdown of lubrication films \cite{Catherall}, can defer the onset of jamming and make it sudden \cite{melrose2}.

Two recent papers \cite{seto,fernandez} support a growing consensus that DST involves a stress-induced transition from lubrication to frictional contacts. Both argue that at low strain rates particles do not touch: contacts  are lubricated, so their static friction coefficient, $m$, is irrelevant. The viscosity of such non-frictional particles would diverge only at random close packing, $\phi_{0}\approx 0.64$ \cite{olsson,lernerpnas}. The repulsive interaction that prevents contact formation is overcome at large stress \cite{seto}, converting the system into an immersed  assembly of frictional grains, whose viscosity diverges instead at $\phi_m <\phi_0$. CST is then argued to arise, for $\phi<\phi_m$, by a stress-induced crossover from the moderate viscosity of the lubricated state to the much higher one of the frictional contact network. 
(Inertia, although sometimes present \cite{fernandez}, is inessential to this basic argument.)
For $\phi>\phi_m$ the system is completely jammed at high stress. Only in this regime is any form of DST predicted in \cite{seto,fernandez}.

This mechanism offers a promising explanation of several key observations: emulsion droplets don't shear thicken as 
friction is
never present; strongly attractive colloids 
generally don't either, as it is
never absent. The onset stress scale $P^*$ for thickening varies strongly with interactions but only weakly with $\phi$.

These successes make it crucial to learn whether other features, such as flow-curve hysteresis, can be explained within the same framework. Also it is by no means clear that the post-DST state is always fully jammed (flowable only by fracture or particle deformation \cite{seto,brown3}). Instead, homogeneously flowing states for rigid particles at $\phi<\phi_m$ could be governed by an S-shaped flow curve $\sigma(\dot\gamma)$, allowing hysteretic DST between a lower (lubricated) and upper (frictional) branch at equal strain rate. Such flow curves, as were seen in a model for Brownian colloids that treats DST as a stress-induced glass transition \cite{holmes1,holmes2}, can also explain the capillarity-induced bistability of a millimetre-scale granule between solid and flowable states \cite{cateshaw}. Such granulation phenomena are poorly understood and industrially important \cite{cateshaw}. 

In this Letter we establish that S-shaped flow curves do arise generically within the scenario of stress-induced contact proliferation. Our work explains hysteresis of DST flow curves, and shows DST to arise for $\phi<\phi_m$, creating an important distinction between DST and complete jamming \cite{seto,fernandez,brown3}. Also, our findings immediately generalize the analysis of granulation in \cite{cateshaw} from the academic case of Brownian frictionless colloids to the industrial mainstream of frictional non-Brownian suspensions.

{\em Phenomenological Analysis:} We consider  a solvent of viscosity $\eta_0$ containing (non-Brownian, density-matched, non-inertial) hard frictional particles that interact with an additional, finite repulsive force, whose range is very small compared to the particle radius $R$. Its strength sets a characteristic scale $P^*$ of particle pressure $P$ that the system can sustain without making frictional contacts \cite{brown3}.  For $p\equiv P/P^* \ll 1$, particles behave as though frictionless, whereas for $p\gg 1$ one recovers an assembly of frictional grains. 
Note that for a fixed maximum force $F^*$ between particles, $P^*\sim F^*R^{-2}$ \cite{barnes}.

We first assume that the jamming density $\phi_m$ (which may depend on the microscopic friction coefficient $m$ between grains) is known. Without repulsions $p = \infty$, and the single parameter theory of \cite{boyer} is recovered. This states that with  $I_v = \eta_0\dot\gamma /P$
\begin{equation}
\phi = \Phi_r(I_v) \;\;\;;\;\;\; \sigma/P = \mu_r(I_v) \label{boyercon}
\end{equation}
where subscript $r$ denotes frictional or `rough' particles. Functions $\Phi_r$ and $\mu_r$ were measured in \cite{boyer} using a semipermeable rheometer at controlled $P$. At fixed $\phi$ these constitutive laws imply quasi-Newtonian scalings,  $P,\sigma \propto \eta_r(\phi)\dot\gamma$. The suspension viscosity $\eta_r(\phi)$ diverges like $(\phi_{m}-\phi)^{-\beta_r}$, where $\beta_r\simeq 2$; in contrast, $\sigma/P$, which like $\eta_r$ is a function of $\phi$ only, has no divergences {\blue \cite{morrisboulay}}.

A similar but distinct one-parameter theory must also emerge when $P^*\to\infty $ so that lubrication films never break. Given the infinitesimal range of our repulsions, this limit should also describe the physics of frictionless or `smooth' grains, $m = 0$. Thus for this `smooth' case
\begin{equation}
\phi = \Phi_{s}(I_v) \;\;\;;\;\;\; \sigma/P = \mu_{s}(I_v) \label{smoothcon}
\end{equation}
Once again  $P,\sigma \propto \eta_{s}(\phi)\dot\gamma$, but $\eta_{s}(\phi)$ now diverges as $(\phi_{0}-\phi)^{-\beta_{s}}$ with $\phi_{0}\simeq \phi_{RCP} \simeq 0.64$. The exponent $\beta_{s}$ is 
similar to $\beta_{r} = 2$, while $\mu_s(I_v)$ has broadly similar properties to $\mu_r(I_v)$ \cite{lernerpnas,thesepeyneau}.

A crossover between the competing divergences of $\eta_s(\phi)$ and $\eta_r(\phi)$, controlled by $p$, can explain both CST for $\phi<\phi_{m}$ \cite{seto} and complete jamming (identified with DST in \cite{seto,fernandez}) for $\phi>\phi_{m}$. We next establish that a smooth interpolation between these divergences also implies sigmoidal, hysteretic flow curves (causing DST between two flowing states of finite viscosity) for  $\phi<\phi_{m}$.

For simplicity we assume $\beta_{s,r} = 2$, and interpolate the divergences with a $p$-dependent jamming density $\phi_J(p)$:
\begin{eqnarray}
P &=& \lambda\dot\gamma (\phi_J(p)-\phi)^{-2} \label{Pform}\\
\phi_J(p) &=& \phi_{m}f+\phi_{0}(1-f) \label{phi_Jform}
\end{eqnarray}
Here $\lambda$ is a constant and $f(p)\in[0,1]$ is the fraction of lubrication films that have ruptured to form frictional contacts.
This creates a two-parameter model
\begin{equation}
\phi = \Phi(I_v,p) \;\;\;;\;\;\; \sigma/P = \mu(I_v,p) \label{twopar}
\end{equation}
where (\ref{Pform},\ref{phi_Jform}) fix the singular terms in  $\Phi(I_v,p)$. (Smooth terms 
{\blue could be added to give}
realistic behavior at small $\phi$ \cite{boyer}.) 
{\blue Further inputs are needed to find} 
the stress ratio $\mu(I_v,p)$. However, since its limiting forms at $p = 0,\infty$ (i.e., $\mu_{r,s}(I_v)$) are comparable nonsingular functions of $\phi$ only, we first assume that  $\mu  = \mu(\phi)$ (but later identify some effects of $p$-dependence). 
  Thus at any fixed $\phi$ the flow curve $\sigma(\dot\gamma)$ has exactly the same shape as $P(\dot\gamma)$.

Sigmoidal flow curves then arise, in a window of densities just below $\phi_{m}$, unless $f(p)$ represents an \textit{unusually slow crossover}. To understand this, consider the ratio $\dot\gamma(P)/P$, which must decrease by a large factor on raising $P$. To achieve that while keeping $d\dot\gamma/dP\ge 0$ in fact requires 
$1-f(p) \ge {\cal O}(p^{-1/2})$ at large $p$. (This can be found by setting $\phi = \phi_{m}$ and requiring $d\dot\gamma/dP\to 0^+$, which imposes a near-vertical but single-valued flow curve.)

{\em Microscopics:} We next give a microscopic discussion of contact network evolution that supports the form (\ref{phi_Jform}) for the crossover. For frictionless particles the coordination of the network of contacts $z$ approaches a critical value $z_c$ at jamming \cite{lernerpnas}. Per particle, the number of soft modes -- collective displacement of particles that do not generate overlaps -- vanishes with $\delta z\equiv z_c-z$, where $z_c=2d$ in $d$ dimensions. This causes the viscosity to diverge as \cite{Lerner2012}
\be
\label{2}
P = A_0 \eta_0\dot\gamma \delta z^{-\alpha}
\ee
where $A_0$ is a constant and $\alpha\simeq 2.85$. In frictional packings, counting soft modes is slightly more involved; nonetheless these must be present for a system of hard particles to flow, and it is found numerically that at the critical state $\phi_{m}$ the number of soft modes is just zero \cite{Kruyt}. Both facts suggest that the loss of soft modes again causes the viscosity divergence. We shall thus assume that Eq.(\ref{2}) is valid both for all packings, so long as $\delta z$ represents the actual number of soft modes per particle. Theoretically the dependence of $\delta z$ on $\phi$ is not derived, but follows empirically from the observed divergences for `rough' and `smooth' particles (with constants $A_{r,s}$) as
\ba
\label{3}
\delta z_r&=&A_r(\phi_{m}-\phi)^{\beta_r/\alpha}\\
\label{4}
\delta z_s&=&A_s(\phi_{0}-\phi)^{\beta_s/\alpha}
\ea

\begin{figure} [ht]
{\includegraphics[width=.9\linewidth]{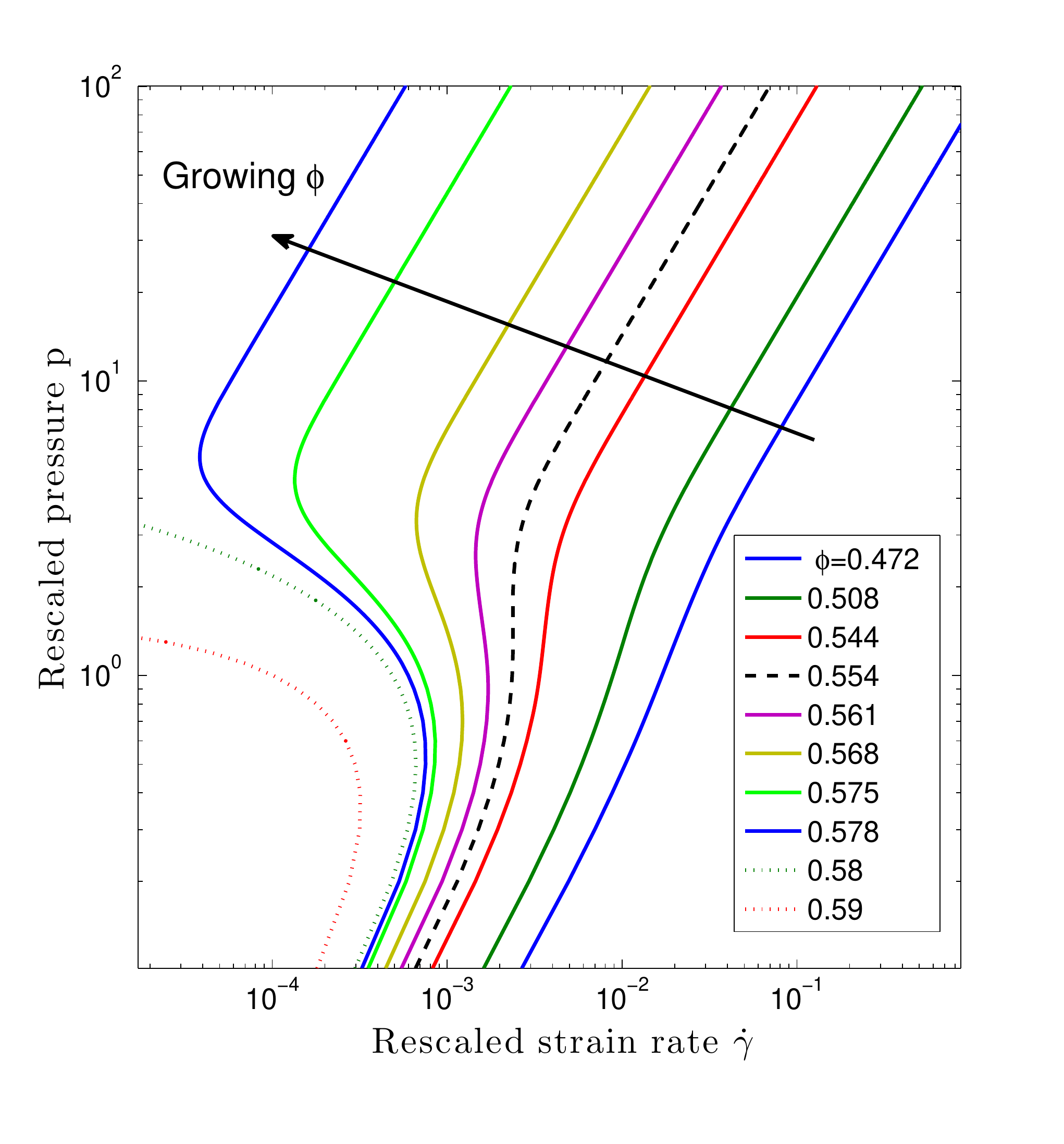}}\\
\caption{Log-log plot of flow curve $p(\dot\gamma)$ from (\ref{Pform},\ref{phi_Jform}) with $\lambda = 1$, $\phi_{0} = 0.64$ and $\phi_{m}=0.58$, for various $\phi$. For small $\phi$, the behavior is near-Newtonian. As $\phi$ increases, CST becomes pronounced; its onset pressure $p\simeq 1$ barely depends on $\phi$ (unlike the corresponding strain rate). The dashed line is for $\phi = \phi_{DST}$. For $\phi_{DST}<\phi<\phi_{m}$, DST is predicted with hysteresis between two flowing, unjammed states. For $\phi>\phi_{m}$ (dotted lines) homogeneous flow can only occur at small strain rates. 
 }
\label{f1}
\end{figure}

Any given packing has a definite $z$; but the number of soft modes $\delta z$ depends on the fraction $f(p)$ of frictional contacts. 
The problem of counting soft modes is somewhat subtle for spherical particles, but we expect 
the rheology of spherical and aspherical particles to display only minor differences \footnote[1]{For spherical grains, the additional constraints at first mainly eliminate the rotational zero-modes of the particles. This effect gives an initially sublinear change in $\delta z$ as $f$ increases (with linearity restored at large $f$), leading to slight changes in the phase diagram of Fig.(\ref{2}). In particular, the maximal stress for a flowing state does not vanish smoothly at $\phi_0$ as shown in the inset.}. For aspherical grains the number of soft modes simply decreases as the number of constraints increases. The latter should increase linearly with the number of frictional contacts, leading to:
\be
\label{5}
\delta z= f(p) \delta z_r +(1-f(p))\delta z_s
\ee
Eqs.(\ref{2}-\ref{5}) are closed. For simplicity we assume (in qualitative accord with the empirical results) that $A_r = A_s = A$, and $\alpha =\beta_r=\beta_s = 2$.
This gives results completely equivalent to (\ref{Pform},\ref{phi_Jform}), with $\lambda = A_0\eta_0/A^2$. (From now on we choose rescaled units where $\lambda =1$.) As already made clear, details of the crossover function $f(p)$ are unimportant unless its decay to unity at large $p$ is very slow. 

{\em Results and Discussion:} We next present numerical results for a suitably bland choice, $f(p)=1-\exp(-p)$. 
The resulting flow curves $P(\dot\gamma)$ are shown in Fig.(\ref{f1}). A key finding is the onset of DST at a packing fraction $\phi_{DST}\approx 0.55$, distinctly below $\phi_{m} = 0.58$.
As $\phi$ approaches $\phi_{DST}$ from below, the slope of the flow curves become more and more pronounced for $p\sim 1$, implying a growing CST. 
In our model, which neglects inertia, at higher $\dot\gamma$ this crosses over to a second Newtonian regime of high viscosity.
At  $\phi_{DST}$ the slope is vertical, and for $\phi_{DST}<\phi<\phi_{m}$, the flow curve is sigmoidal, signaling hysteretic DST between upper and lower branches of finite viscosity. The maximal extent of hysteresis is delineated by two strain rates $\dot\gamma^+>\dot \gamma^-$ where $d\dot\gamma/dP=0$.  For $\phi \to \phi_{m}$, we find $\dot\gamma^- \to 0$. At this point, the upper branch of the sigmoid disappears, signifying complete jamming. 
For $\phi\ge\phi_{m}$ material is flowable at low stress, but completely jammed for $p\gg 1$. One may still observe a discontinuous (and possibly hysteretic) thickening at $\dot\gamma^+$, but the thickened state must flow inhomogeneously. 

\begin{figure}[!hbt]
  {\includegraphics[width=1.0\linewidth]{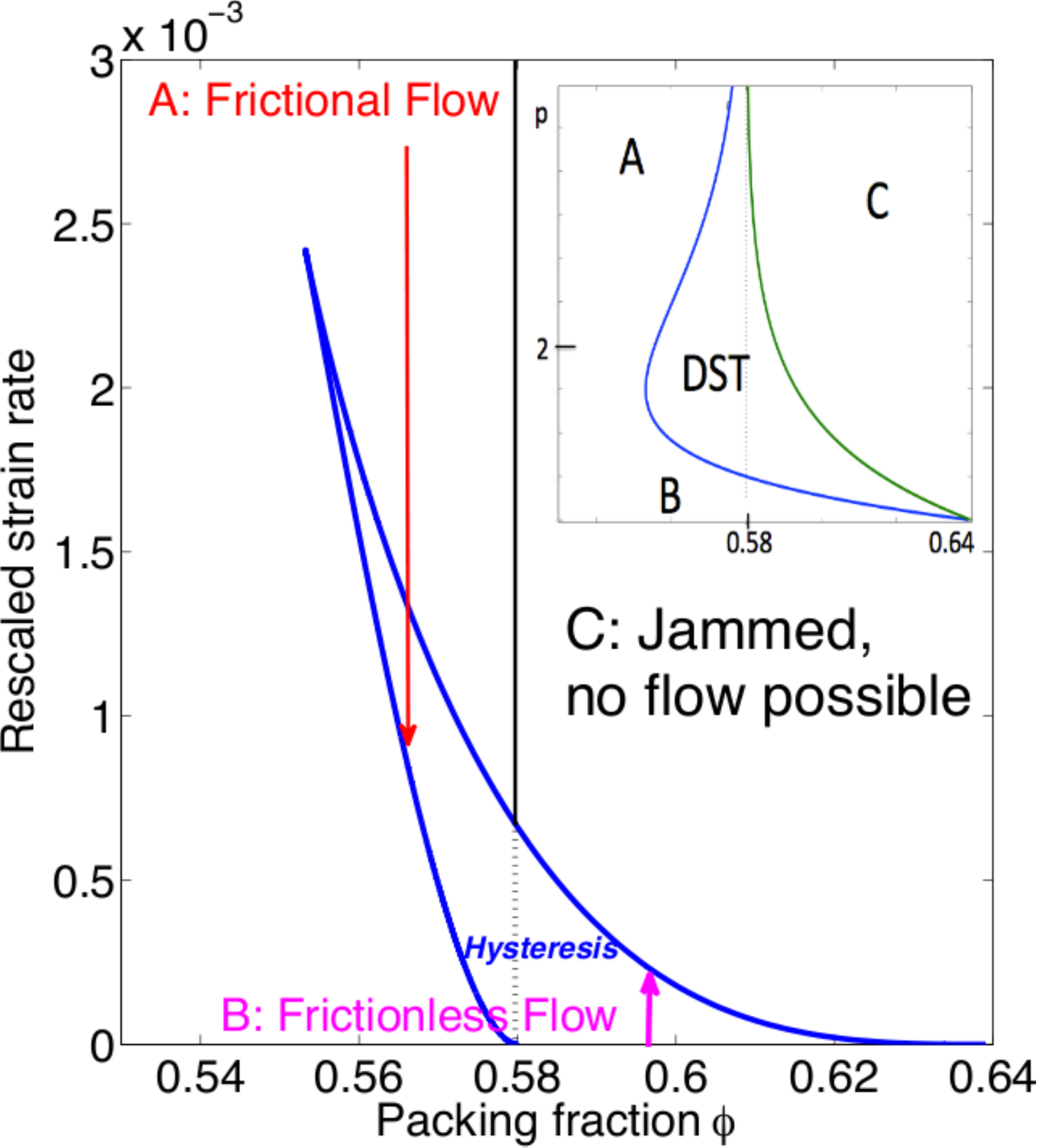}}
\caption{Phase diagram in the $(\phi,\dot\gamma)$ plane. The solid (blue) curves delimit upper and lower stability limits $\dot\gamma^\pm(\phi)$ between which hysteresis is possible. The cusp at the left of this region identifies $\phi = \phi_{DST}$. For $\phi_0>\phi > \phi_{DST}$, increasing the strain rate gives an upward stress-jump at the upper stability curve: ascending (pink) arrow. 
To the right of the vertical line at $\phi = \phi_{m}$,
homogeneous flow is impossible for $\dot\gamma >\dot\gamma^+(\phi)$ unless particles are deformable. For $\phi<\phi_{m}$, the upward jump is to a flowing frictional state; lowering the strain rate from a large value gives a negative stress jump at the lower stability limit $\dot\gamma^-(\phi)$: descending (red) arrow. Inset: Phase diagram in the $(p,\phi)$ plane. The left (blue) curve locates the stress-jump on steadily increasing strain rate.
The right (green) curve shows the maximal stress above which no flowing states exist.}
\label{f2}
\end{figure}

Fig.(\ref{f2}) shows a phase diagram of the various flow regimes. Inside the solid (blue) curve, there is hysteresis and flow can depend on strain-rate history. Several features of this diagram do not depend on the details of $f$: (a) near  $\phi_{DST}$ the hysteresis zone narrows to a cusp, with $\dot\gamma^+-\dot\gamma^-\propto (\phi-\phi_{DST})^{3/2}$, as expected from a saddle node bifurcation; (b) on the approach to complete jamming, $\dot\gamma^-$ vanishes at least as $(\phi_{m}-\phi)^2$, and for $f'(0)>0$ as $(\phi_{m}-\phi)^3$ (modulo logarithmic corrections); (c) $\dot \gamma^+$ vanishes only at $\phi_{0}$ beyond which homogeneous flow is impossible even at infinitesimal $\dot\gamma$.

In the presence of noise, jumps can occur {\em before} the relevant stability limit is 
reached: the hysteretic
regime in Fig.(\ref
{f2}) represents the maximum possible. 
(Noise-induced nucleation might recover a single-valued but discontinuous curve as $d\dot\gamma/dt\to 0$, but this limit could in turn prove experimentally inaccessible \cite{grand}.)
Note also that at DST, where $I_v$ jumps downward and $p$ up, one expects a jump in the stress ratio $\mu = \sigma/P$ which depends in on the full form of $\mu(I_v,p)$. (However numerics support that $\mu$ weakly depends on friction at fixed $\phi$ \cite{trulsson}, so this effect may be small). The same applies to other stress ratios, such as those involving normal stress differences.

In the inset of  Fig.(\ref{f2}) the same phase diagram is plotted in the $\phi,p$ plane. This might be relevant for experiments at controlled $P$ \cite{boyer}. In principle these might allow one to reach states inaccessible by any flow history at fixed $\phi$, such as those on the decreasing `middle' branch of the flow curve $P(\dot\gamma)$. (For $\phi >\phi_m$ this becomes the upper branch, but is still decreasing.)
However, the same branch is also present for $\sigma(\dot\gamma)$ where its observation at fixed $\sigma$ is normally precluded by 
{\blue transverse shear banding} \cite{olmsted}. Ignoring particle migration (which is slow \cite{besseling}) such banding might be prevented if $P$ is controlled locally (not just as an average along the velocity gradient direction). It is unclear to us whether the semipermeable rheometer of \cite{boyer} achieves this.

Finally we address the role played by the static friction coefficient $m$ of contacts. So long as this is a positive constant, $\phi_m<\phi_0$ and our model remains applicable; both CST and DST are predicted. Since the jamming density $\phi_m$ moves away from $\phi_0$ as $m$ is increased \cite{dacruz}, our model predicts shear thickening to be more pronounced with high friction particles than low ones -- as reported experimentally \cite{fernandez}. However an alternative but similar scenario might now be obtained \textit{even without finite repulsions}, by choosing a stress-dependent contact friction $m(p)$ that  increases with $p = P/P^*$.

{\it Conclusion}: We have provided a phenomenological model of shear thickening for frictional hard spheres with finite  short-range repulsions. Our analysis explains observations of hysteresis, and predicts that DST should begin at an onset packing fraction, $\phi_{DST}<\phi_m$, below the jamming point. Our result may be tested by careful experiments on hysteresis (which should reveal DST to smoothly flowing states) in a system of sufficiently hard particles, at fixed volume fraction. DST (and indeed CST) should disappear altogether if flow is measured at fixed particle pressure $P$ \cite{boyer}. DST also will not be observable if the onset stress $P^*$ exceeds the threshold $\tau/R$ for containment of particles by rheometer menisci of surface tension $\tau$ \cite{brown3,cateshaw}. In this sense DST depends on boundary conditions as well as bulk properties \cite{brown3}; but for $\phi<\phi_{m}$, with fixed $\phi$ and no free surfaces, it reverts to an intrinsic property of the bulk flow curve $\sigma(\dot\gamma)$. We have neglected gravity, Brownian motion, and inertia, thus showing these not to be prerequisites for shear thickening, but it would be interesting to see how much they change the picture. For instance, it may be that slight Brownian motion has effects very similar to a short-range repulsion \cite{melrose}.
Also it is possible that shear thickening by a related but inertial mechanism \cite{fernandez} would arise in fast enough flows even for purely hard spheres, whereas additional short-range repulsions introduce a second, non-inertial mechanism operative at lower strain rates.

Acknowledgments: 
We thank 
Bruno Andreotti,
Paul Chaikin
Eric DeGiuli,
Nicolas Fernandez,
Ben Guy,
Gary Hunter,
Lucio Isa,
Edan Lerner, 
Jie Lin, 
Jeff Morris,
Wilson Poon
and
Le Yan   
for discussions. 
MW thanks NSF CBET Grant 1236378, NSF DMR Grant 1105387, and MRSEC Program of the NSF DMR-0820341 for partial funding. MEC thanks EPSRC J/007404 and the Royal Society for funding, and NYU for hospitality.


\begin{thebibliography}{1}%
\makeatletter
\providecommand \@ifxundefined [1]{%
 \@ifx{#1\undefined}
}%
\providecommand \@ifnum [1]{%
 \ifnum #1\expandafter \@firstoftwo
 \else \expandafter \@secondoftwo
 \fi
}%
\providecommand \@ifx [1]{%
 \ifx #1\expandafter \@firstoftwo
 \else \expandafter \@secondoftwo
 \fi
}%
\providecommand \natexlab [1]{#1}%
\providecommand \enquote  [1]{``#1''}%
\providecommand \bibnamefont  [1]{#1}%
\providecommand \bibfnamefont [1]{#1}%
\providecommand \citenamefont [1]{#1}%
\providecommand \href@noop [0]{\@secondoftwo}%
\providecommand \href [0]{\begingroup \@sanitize@url \@href}%
\providecommand \@href[1]{\@@startlink{#1}\@@href}%
\providecommand \@@href[1]{\endgroup#1\@@endlink}%
\providecommand \@sanitize@url [0]{\catcode `\\12\catcode `\$12\catcode
  `\&12\catcode `\#12\catcode `\^12\catcode `\_12\catcode `\%12\relax}%
\providecommand \@@startlink[1]{}%
\providecommand \@@endlink[0]{}%
\providecommand \url  [0]{\begingroup\@sanitize@url \@url }%
\providecommand \@url [1]{\endgroup\@href {#1}{\urlprefix }}%
\providecommand \urlprefix  [0]{URL }%
\providecommand \Eprint [0]{\href }%
\@ifxundefined \urlstyle {%
  \providecommand \doi  [0]{\begingroup \@sanitize@url \@doi}%
  \providecommand \@doi [1]{\endgroup \@@startlink {\doibase
  #1}doi:\discretionary {}{}{}#1\@@endlink }%
}{%
  \providecommand \doi  [0]{doi:\discretionary{}{}{}\begingroup
  \urlstyle{rm}\Url }%
}%
\providecommand \doibase [0]{http://dx.doi.org/}%
\providecommand \Doi [0]{\begingroup \@sanitize@url \@Doi }%
\providecommand \@Doi  [1]{\endgroup\@@startlink{\doibase#1}\@@Doi}%
\providecommand \@@Doi [1]{#1\@@endlink}%
\providecommand \selectlanguage [0]{\@gobble}%
\providecommand \bibinfo  [0]{\@secondoftwo}%
\providecommand \bibfield  [0]{\@secondoftwo}%
\providecommand \translation [1]{[#1]}%
\providecommand \BibitemOpen [0]{}%
\providecommand \bibitemStop [0]{}%
\providecommand \bibitemNoStop [0]{.\EOS\space}%
\providecommand \EOS [0]{\spacefactor3000\relax}%
\providecommand \BibitemShut  [1]{\csname bibitem#1\endcsname}%
\bibitem [{Note1()}]{Note1}%
  \BibitemOpen
  \bibinfo {note} {For spherical grains, the additional constraints at first
  mainly eliminate the rotational zero-modes of the particles. This effect
  gives an initially sublinear change in $\delta z$ as $f$ increases (with
  linearity restored at large $f$), leading to slight changes in the phase
  diagram of Fig.(\ref {2}). In particular, the maximal stress for a flowing
  state does not vanish smoothly at $\phi _0$ as shown in the
  inset.}\BibitemShut {Stop}%
\end{thebibliography}%


\begin{thebibliography}{31}%
\makeatletter
\providecommand \@ifxundefined [1]{%
 \@ifx{#1\undefined}
}%
\providecommand \@ifnum [1]{%
 \ifnum #1\expandafter \@firstoftwo
 \else \expandafter \@secondoftwo
 \fi
}%
\providecommand \@ifx [1]{%
 \ifx #1\expandafter \@firstoftwo
 \else \expandafter \@secondoftwo
 \fi
}%
\providecommand \natexlab [1]{#1}%
\providecommand \enquote  [1]{``#1''}%
\providecommand \bibnamefont  [1]{#1}%
\providecommand \bibfnamefont [1]{#1}%
\providecommand \citenamefont [1]{#1}%
\providecommand \href@noop [0]{\@secondoftwo}%
\providecommand \href [0]{\begingroup \@sanitize@url \@href}%
\providecommand \@href[1]{\@@startlink{#1}\@@href}%
\providecommand \@@href[1]{\endgroup#1\@@endlink}%
\providecommand \@sanitize@url [0]{\catcode `\\12\catcode `\$12\catcode
  `\&12\catcode `\#12\catcode `\^12\catcode `\_12\catcode `\%12\relax}%
\providecommand \@@startlink[1]{}%
\providecommand \@@endlink[0]{}%
\providecommand \url  [0]{\begingroup\@sanitize@url \@url }%
\providecommand \@url [1]{\endgroup\@href {#1}{\urlprefix }}%
\providecommand \urlprefix  [0]{URL }%
\providecommand \Eprint [0]{\href }%
\providecommand \doibase [0]{http://dx.doi.org/}%
\providecommand \selectlanguage [0]{\@gobble}%
\providecommand \bibinfo  [0]{\@secondoftwo}%
\providecommand \bibfield  [0]{\@secondoftwo}%
\providecommand \translation [1]{[#1]}%
\providecommand \BibitemOpen [0]{}%
\providecommand \bibitemStop [0]{}%
\providecommand \bibitemNoStop [0]{.\EOS\space}%
\providecommand \EOS [0]{\spacefactor3000\relax}%
\providecommand \BibitemShut  [1]{\csname bibitem#1\endcsname}%
\let\auto@bib@innerbib\@empty
\bibitem [{\citenamefont {{Brown}}\ and\ \citenamefont
  {{Jaeger}}(2013)}]{brown3}%
  \bibfield  {author} {\bibinfo {author} {\bibfnamefont {E.}~\bibnamefont
  {{Brown}}}\ and\ \bibinfo {author} {\bibfnamefont {H.~M.}\ \bibnamefont
  {{Jaeger}}},\ }\href@noop {} {\bibfield  {journal} {\bibinfo  {journal}
  {e-print}\ } (\bibinfo {year} {2013})},\ \Eprint
  {http://arxiv.org/abs/1307.0269} {arXiv:1307.0269}
\bibitem [{\citenamefont {Barnes}(1989)}]{barnes}%
  \bibfield  {author} {\bibinfo {author} {\bibfnamefont {H.~A.}\ \bibnamefont
  {Barnes}},\ }\href {\doibase 10.1122/1.550017} {\bibfield  {journal}
  {\bibinfo  {journal} {J. Rheol.}\ }\textbf {\bibinfo {volume}
  {33}},\ \bibinfo {pages} {329} (\bibinfo {year} {1989})}
\bibitem [{\citenamefont {Coussot}\ and\ \citenamefont
  {Wagner}(2009)}]{wagner}%
  \bibfield  {author} {\bibinfo {author} {\bibfnamefont {P.}~\bibnamefont
  {Coussot}}\ and\ \bibinfo {author} {\bibfnamefont {N.}~\bibnamefont
  {Wagner}},\ }\href {http://dx.doi.org/10.1007/s00397-009-0380-x} {\bibfield
  {journal} {\bibinfo  {journal} {Rheologica Acta}\ }\textbf {\bibinfo {volume}
  {48}},\ \bibinfo {pages} {827} (\bibinfo {year} {2009})}
\bibitem [{\citenamefont {Fernandez}\ \emph {et~al.}(2013)\citenamefont
  {Fernandez}, \citenamefont {Mani}, \citenamefont {Rinaldi}, \citenamefont
  {Kadau}, \citenamefont {Mosquet}, \citenamefont {Lombois-Burger},
  \citenamefont {Cayer-Barrioz}, \citenamefont {Herrmann}, \citenamefont
  {Spencer},\ and\ \citenamefont {Isa}}]{fernandez}%
  \bibfield  {author} {\bibinfo {author} {\bibfnamefont {N.}~\bibnamefont
  {Fernandez}}, \bibinfo {author} {\bibfnamefont {R.}~\bibnamefont {Mani}},
  \bibinfo {author} {\bibfnamefont {D.}~\bibnamefont {Rinaldi}}, \bibinfo
  {author} {\bibfnamefont {D.}~\bibnamefont {Kadau}}, \bibinfo {author}
  {\bibfnamefont {M.}~\bibnamefont {Mosquet}}, \bibinfo {author} {\bibfnamefont
  {H.}~\bibnamefont {Lombois-Burger}}, \bibinfo {author} {\bibfnamefont
  {J.}~\bibnamefont {Cayer-Barrioz}}, \bibinfo {author} {\bibfnamefont {H.~J.}\
  \bibnamefont {Herrmann}}, \bibinfo {author} {\bibfnamefont {N.~D.}\
  \bibnamefont {Spencer}}, \ and\ \bibinfo {author} {\bibfnamefont
  {L.}~\bibnamefont {Isa}},\ }\href {\doibase 10.1103/PhysRevLett.111.108301}
  {\bibfield  {journal} {\bibinfo  {journal} {Phys. Rev. Lett.}\ }\textbf
  {\bibinfo {volume} {111}},\ \bibinfo {pages} {108301} (\bibinfo {year}
  {2013})}
\bibitem [{\citenamefont {Fall}\ \emph {et~al.}(2010)\citenamefont {Fall},
  \citenamefont {Lemaitre}, \citenamefont {Bertrand}, \citenamefont {Bonn},\
  and\ \citenamefont {Ovarlez}}]{fall}%
  \bibfield  {author} {\bibinfo {author} {\bibfnamefont {A.}~\bibnamefont
  {Fall}}, \bibinfo {author} {\bibfnamefont {A.}~\bibnamefont {Lemaitre}},
  \bibinfo {author} {\bibfnamefont {F.}~\bibnamefont {Bertrand}}, \bibinfo
  {author} {\bibfnamefont {D.}~\bibnamefont {Bonn}}, \ and\ \bibinfo {author}
  {\bibfnamefont {G.}~\bibnamefont {Ovarlez}},\ }\href {\doibase
  10.1103/PhysRevLett.105.268303} {\bibfield  {journal} {\bibinfo  {journal}
  {Phys. Rev. Lett.}\ }\textbf {\bibinfo {volume} {105}},\ \bibinfo {pages}
  {268303} (\bibinfo {year} {2010})}
\bibitem [{\citenamefont {Cheng}\ \emph {et~al.}(2011)\citenamefont {Cheng},
  \citenamefont {McCoy}, \citenamefont {Israelachvili},\ and\ \citenamefont
  {Cohen}}]{Cheng}%
  \bibfield  {author} {\bibinfo {author} {\bibfnamefont {X.}~\bibnamefont
  {Cheng}}, \bibinfo {author} {\bibfnamefont {J.~H.}\ \bibnamefont {McCoy}},
  \bibinfo {author} {\bibfnamefont {J.~N.}\ \bibnamefont {Israelachvili}}, \
  and\ \bibinfo {author} {\bibfnamefont {I.}~\bibnamefont {Cohen}},\ }\href
  {\doibase 10.1126/science.1207032} {\bibfield  {journal} {\bibinfo  {journal}
  {Science}\ }\textbf {\bibinfo {volume} {333}},\ \bibinfo {pages} {1276}
  (\bibinfo {year} {2011})}
\bibitem [{\citenamefont {Farr}\ \emph {et~al.}(1997)\citenamefont {Farr},
  \citenamefont {Melrose},\ and\ \citenamefont {Ball}}]{farr}%
  \bibfield  {author} {\bibinfo {author} {\bibfnamefont {R. S.}~\bibnamefont
  {Farr}}, \bibinfo {author} {\bibfnamefont {J.~R.}\ \bibnamefont {Melrose}}, \
  and\ \bibinfo {author} {\bibfnamefont {R. C.}~\bibnamefont {Ball}},\ }\href@noop
  {} {\bibfield  {journal} {\bibinfo  {journal} {Phys. Rev. E}\ }\textbf
  {\bibinfo {volume} {55}},\ \bibinfo {pages} {7203} (\bibinfo {year}
  {1997})}
\bibitem [{\citenamefont {Brady}\ and\ \citenamefont {Bossis}(1988)}]{brady2}%
  \bibfield  {author} {\bibinfo {author} {\bibfnamefont {J.~F.}\ \bibnamefont
  {Brady}}\ and\ \bibinfo {author} {\bibfnamefont {G.}~\bibnamefont {Bossis}},\
  }\href@noop {} {\bibfield  {journal} {\bibinfo  {journal} {Annual Rev. 
  Fluid Mech.}\ }\textbf {\bibinfo {volume} {20}},\ \bibinfo {pages} {111}
  (\bibinfo {year} {1988})}
\bibitem [{\citenamefont {Hoffman}(1974)}]{Hoffman}%
  \bibfield  {author} {\bibinfo {author} {\bibfnamefont {R.}~\bibnamefont
  {Hoffman}},\ }\href@noop {} {\bibfield  {journal} {\bibinfo  {journal}
  {J. Colloid Interface Sci.}\ }\textbf {\bibinfo {volume}
  {46}},\ \bibinfo {pages} {491} (\bibinfo {year} {1974})}
\bibitem [{\citenamefont {Brown}\ and\ \citenamefont {Jaeger}(2009)}]{brown1}%
  \bibfield  {author} {\bibinfo {author} {\bibfnamefont {E.}~\bibnamefont
  {Brown}}\ and\ \bibinfo {author} {\bibfnamefont {H.~M.}\ \bibnamefont
  {Jaeger}},\ }\href@noop {} {\bibfield  {journal} {\bibinfo  {journal} {Phys.
  Rev. Lett.}\ }\textbf {\bibinfo {volume} {103}},\ \bibinfo {pages} {086001}
  (\bibinfo {year} {2009})}
\bibitem [{\citenamefont {Brown}\ \emph {et~al.}(2010)\citenamefont {Brown},
  \citenamefont {Forman}, \citenamefont {Orellana}, \citenamefont {Zhang},
  \citenamefont {Maynor}, \citenamefont {Betts}, \citenamefont {DeSimone},\
  and\ \citenamefont {Jaeger}}]{brown2}%
  \bibfield  {author} {\bibinfo {author} {\bibfnamefont {E.}~\bibnamefont
  {Brown}}, \bibinfo {author} {\bibfnamefont {N.~A.}\ \bibnamefont {Forman}},
  \bibinfo {author} {\bibfnamefont {C.~S.}\ \bibnamefont {Orellana}}, \bibinfo
  {author} {\bibfnamefont {H.}~\bibnamefont {Zhang}}, \bibinfo {author}
  {\bibfnamefont {B.~W.}\ \bibnamefont {Maynor}}, \bibinfo {author}
  {\bibfnamefont {D.~E.}\ \bibnamefont {Betts}}, \bibinfo {author}
  {\bibfnamefont {J.~M.}\ \bibnamefont {DeSimone}}, \ and\ \bibinfo {author}
  {\bibfnamefont {H.~M.}\ \bibnamefont {Jaeger}},\ }\href {\doibase
  10.1038/nmat2627} {\bibfield  {journal} {\bibinfo  {journal} {Nat. Mater.}\
  }\textbf {\bibinfo {volume} {9}},\ \bibinfo {pages} {220} (\bibinfo {year}
  {2010})}
\bibitem [{\citenamefont {Boyer}\ \emph {et~al.}(2011)\citenamefont {Boyer},
  \citenamefont {Guazzelli},\ and\ \citenamefont {Pouliquen}}]{boyer}%
  \bibfield  {author} {\bibinfo {author} {\bibfnamefont {F.}~\bibnamefont
  {Boyer}}, \bibinfo {author} {\bibfnamefont {E.}~\bibnamefont {Guazzelli}}, \
  and\ \bibinfo {author} {\bibfnamefont {O.}~\bibnamefont {Pouliquen}},\ }\href
  {\doibase 10.1103/PhysRevLett.107.188301} {\bibfield  {journal} {\bibinfo
  {journal} {Phys. Rev. Lett.}\ }\textbf {\bibinfo {volume} {107}},\ \bibinfo
  {pages} {188301} (\bibinfo {year} {2011})}
\bibitem [{\citenamefont {Lemaitre}\ \emph {et~al.}(2009)\citenamefont
  {Lemaitre}, \citenamefont {Roux},\ and\ \citenamefont
  {Chevoir}}]{lemaitreroux}%
  \bibfield  {author} {\bibinfo {author} {\bibfnamefont {A.}~\bibnamefont
  {Lemaitre}}, \bibinfo {author} {\bibfnamefont {J.-N.}\ \bibnamefont {Roux}},
  \ and\ \bibinfo {author} {\bibfnamefont {F.}~\bibnamefont {Chevoir}},\
  }\href@noop {} {\bibfield  {journal} {\bibinfo  {journal} {Rheologica Acta}\
  }\textbf {\bibinfo {volume} {48}},\ \bibinfo {pages} {925} (\bibinfo {year}
  {2009})}
\bibitem [{\citenamefont {Cates}\ \emph {et~al.}(2005)\citenamefont {Cates},
  \citenamefont {Haw},\ and\ \citenamefont {Holmes}}]{cateshaw}%
  \bibfield  {author} {\bibinfo {author} {\bibfnamefont {M.}~\bibnamefont
  {Cates}}, \bibinfo {author} {\bibfnamefont {M. E.}~\bibnamefont {Haw}}, \ and\
  \bibinfo {author} {\bibfnamefont {C.}~\bibnamefont {Holmes}},\ }\href@noop {}
  {\bibfield  {journal} {\bibinfo  {journal} {J. Phys. Cond.
  Mat.}\ }\textbf {\bibinfo {volume} {17}},\ \bibinfo {pages} {S2517}
  (\bibinfo {year} {2005})}
\bibitem [{\citenamefont {Besseling}\ \emph {et~al.}(2010)\citenamefont
  {Besseling}, \citenamefont {Isa}, \citenamefont {Ballesta}, \citenamefont
  {Petekidis}, \citenamefont {Cates},\ and\ \citenamefont {Poon}}]{besseling}%
  \bibfield  {author} {\bibinfo {author} {\bibfnamefont {R.}~\bibnamefont
  {Besseling}}, \bibinfo {author} {\bibfnamefont {L.}~\bibnamefont {Isa}},
  \bibinfo {author} {\bibfnamefont {P.}~\bibnamefont {Ballesta}}, \bibinfo
  {author} {\bibfnamefont {G.}~\bibnamefont {Petekidis}}, \bibinfo {author}
  {\bibfnamefont {M. E.}~\bibnamefont {Cates}}, \ and\ \bibinfo {author}
  {\bibfnamefont {W. C. K.}~\bibnamefont {Poon}},\ }\href@noop {} {\bibfield
  {journal} {\bibinfo  {journal} {Phys. Rev. Lett.}\ }\textbf {\bibinfo
  {volume} {105}},\ \bibinfo {pages} {268301} (\bibinfo {year}
  {2010})}
\bibitem [{\citenamefont {Catherall}\ \emph {et~al.}(2000)\citenamefont
  {Catherall}, \citenamefont {Melrose},\ and\ \citenamefont
  {Ball}}]{Catherall}%
  \bibfield  {author} {\bibinfo {author} {\bibfnamefont {A.~A.}\ \bibnamefont
  {Catherall}}, \bibinfo {author} {\bibfnamefont {J.~R.}\ \bibnamefont
  {Melrose}}, \ and\ \bibinfo {author} {\bibfnamefont {R.~C.}\ \bibnamefont
  {Ball}},\ }\href@noop {} {\bibfield  {journal} {\bibinfo  {journal} {J. Rheol.}\ }\textbf {\bibinfo {volume} {44}},\ \bibinfo {pages} {1}
  (\bibinfo {year} {2000})}
\bibitem [{\citenamefont {Melrose}\ and\ \citenamefont
  {Ball}(1995)}]{melrose2}%
  \bibfield  {author} {\bibinfo {author} {\bibfnamefont {J. R.}~\bibnamefont
  {Melrose}}\ and\ \bibinfo {author} {\bibfnamefont {R. C.}~\bibnamefont {Ball}},\
  }\href@noop {} {\bibfield  {journal} {\bibinfo  {journal} { Europhys.
  Lett.}\ }\textbf {\bibinfo {volume} {32}},\ \bibinfo {pages} {535}
  (\bibinfo {year} {1995})}
\bibitem [{\citenamefont {{Seto}}\ \emph {et~al.}(2013)\citenamefont {{Seto}},
  \citenamefont {{Mari}}, \citenamefont {{Morris}},\ and\ \citenamefont
  {{Denn}}}]{seto}%
  \bibfield  {author} {\bibinfo {author} {\bibfnamefont {R.}~\bibnamefont
  {{Seto}}}, \bibinfo {author} {\bibfnamefont {R.}~\bibnamefont {{Mari}}},
  \bibinfo {author} {\bibfnamefont {J.~F.}\ \bibnamefont {{Morris}}}, \ and\
  \bibinfo {author} {\bibfnamefont {M.~M.}\ \bibnamefont {{Denn}}},\
  }\href@noop {} {\bibfield  {journal} {\bibinfo  {journal} {e-print}\ }
  (\bibinfo {year} {2013})},\ \Eprint {http://arxiv.org/abs/1306.5985}
  {arXiv:1306.5985} 
\bibitem [{\citenamefont {{Olsson}}\ and\ \citenamefont
  {{Teitel}}(2007)}]{olsson}%
  \bibfield  {author} {\bibinfo {author} {\bibfnamefont {P.}~\bibnamefont
  {{Olsson}}}\ and\ \bibinfo {author} {\bibfnamefont {S.}~\bibnamefont
  {{Teitel}}},\ }\href@noop {} {\bibfield  {journal} {\bibinfo  {journal}
  {Phys.\ Rev.\ Lett.}\ }\textbf {\bibinfo {volume} {99}},\ \bibinfo {pages}
  {178001} (\bibinfo {year} {2007})}
\bibitem [{\citenamefont {Lerner}\ \emph
  {et~al.}(2012{\natexlab{a}})\citenamefont {Lerner}, \citenamefont {DŸring},\
  and\ \citenamefont {Wyart}}]{lernerpnas}%
  \bibfield  {author} {\bibinfo {author} {\bibfnamefont {E.}~\bibnamefont
  {Lerner}}, \bibinfo {author} {\bibfnamefont {G.}~\bibnamefont {DŸring}}, \
  and\ \bibinfo {author} {\bibfnamefont {M.}~\bibnamefont {Wyart}},\ }\href
  {\doibase 10.1073/pnas.1120215109} {\bibfield  {journal} {\bibinfo  {journal}
  {Proc. Nat. Acad. Sci. USA}\ }\textbf {\bibinfo
  {volume} {109}},\ \bibinfo {pages} {4798} (\bibinfo {year}
  {2012}{\natexlab{a}})}
\bibitem [{\citenamefont {Holmes}\ \emph {et~al.}(2005)\citenamefont {Holmes},
  \citenamefont {Cates}, \citenamefont {Fuchs},\ and\ \citenamefont
  {Sollich}}]{holmes1}%
  \bibfield  {author} {\bibinfo {author} {\bibfnamefont {C.~B.}\ \bibnamefont
  {Holmes}}, \bibinfo {author} {\bibfnamefont {M.~E.}\ \bibnamefont {Cates}},
  \bibinfo {author} {\bibfnamefont {M.}~\bibnamefont {Fuchs}}, \ and\ \bibinfo
  {author} {\bibfnamefont {P.}~\bibnamefont {Sollich}},\ }\href@noop {}
  {\bibfield  {journal} {\bibinfo  {journal} {J. Rheol.}\ }\textbf
  {\bibinfo {volume} {49}},\ \bibinfo {pages} {237} (\bibinfo {year}
  {2005})}
\bibitem [{\citenamefont {Holmes}\ \emph {et~al.}(2003)\citenamefont {Holmes},
  \citenamefont {Fuchs},\ and\ \citenamefont {Cates}}]{holmes2}%
  \bibfield  {author} {\bibinfo {author} {\bibfnamefont {C.~B.}\ \bibnamefont
  {Holmes}}, \bibinfo {author} {\bibfnamefont {M.}~\bibnamefont {Fuchs}}, \
  and\ \bibinfo {author} {\bibfnamefont {M.~E.}\ \bibnamefont {Cates}},\
  }\href@noop {} {\bibfield  {journal} {\bibinfo  {journal} {Europhys.
  Lett.}\ }\textbf {\bibinfo {volume} {63}},\ \bibinfo {pages} {240}
  (\bibinfo {year} {2003})}

\bibitem{morrisboulay} {\blue 
J. F. Morris and F. Boulay, J. Rheol. {\bf 43}, 1213 (1999).
}

\bibitem [{\citenamefont {Peyneau}(2009)}]{thesepeyneau}%
  \bibfield  {author} {\bibinfo {author} {\bibfnamefont {P.-E.}\ \bibnamefont
  {Peyneau}},\ }\href@noop {} {\emph {\bibinfo {title} {{Etude du Comportement
  et du Compactage de Pates Granulaires par Simulation Numerique Discrete}}}}\ \href@noop {} {Ph.D. Thesis},\ \bibinfo  {school} {Ecole des
  Ponts ParisTech}
  (\bibinfo {year} {{2009}})
\bibitem [{\citenamefont {Lerner}\ \emph
  {et~al.}(2012{\natexlab{b}})\citenamefont {Lerner}, \citenamefont {During},\
  and\ \citenamefont {Wyart}}]{Lerner2012}%
  \bibfield  {author} {\bibinfo {author} {\bibfnamefont {E.}~\bibnamefont
  {Lerner}}, \bibinfo {author} {\bibfnamefont {G.}~\bibnamefont {During}}, \
  and\ \bibinfo {author} {\bibfnamefont {M.}~\bibnamefont {Wyart}},\ }\href
  {http://stacks.iop.org/0295-5075/99/i=5/a=58003} {\bibfield  {journal}
  {\bibinfo  {journal} {EPL}\ }\textbf {\bibinfo {volume}
  {99}},\ \bibinfo {pages} {58003} (\bibinfo {year}
  {2012}{\natexlab{b}})}
\bibitem [{\citenamefont {Kruyt}(2010)}]{Kruyt}%
  \bibfield  {author} {\bibinfo {author} {\bibfnamefont {N.~P.}\ \bibnamefont
  {Kruyt}},\ }\href@noop {} {\bibfield  {journal} {\bibinfo  {journal} {Comptes
  Rendus Mecanique}\ }\textbf {\bibinfo {volume} {338}},\ \bibinfo {pages} {596
  } (\bibinfo {year} {2010})}
\bibitem [{Note1()}]{Note1}%
  \bibinfo {note} {For spherical grains, the additional constraints at first
  mainly eliminate the rotational zero-modes of the particles. This effect
  gives an initially sublinear change in $\delta z$ as $f$ increases (with
  linearity restored at large $f$), leading to slight changes in the phase
  diagram of Fig.(\ref {2}). In particular, the maximal stress for a flowing
  state does not vanish smoothly at $\phi _0$ as shown in the
  inset.}
\bibitem [{\citenamefont {Grand}\ \emph {et~al.}(1997)\citenamefont {Grand},
  \citenamefont {Arrault},\ and\ \citenamefont {Cates}}]{grand}%
  \bibfield  {author} {\bibinfo {author} {\bibfnamefont {C.}~\bibnamefont
  {Grand}}, \bibinfo {author} {\bibfnamefont {J.}~\bibnamefont {Arrault}}, \
  and\ \bibinfo {author} {\bibfnamefont {M. E.}~\bibnamefont {Cates}},\
  }\href@noop {} {\bibfield  {journal} {\bibinfo  {journal} {J. de
  Physique II}\ }\textbf {\bibinfo {volume} {7}},\ \bibinfo {pages} {1071}
  (\bibinfo {year} {1997})}
\bibitem [{\citenamefont {Trulsson}\ \emph {et~al.}(2012)\citenamefont
  {Trulsson}, \citenamefont {Andreotti},\ and\ \citenamefont
  {Claudin}}]{trulsson}%
  \bibfield  {author} {\bibinfo {author} {\bibfnamefont {M.}~\bibnamefont
  {Trulsson}}, \bibinfo {author} {\bibfnamefont {B.}~\bibnamefont {Andreotti}},
  \ and\ \bibinfo {author} {\bibfnamefont {P.}~\bibnamefont {Claudin}},\
  }\href@noop {} {\bibfield  {journal} {\bibinfo  {journal} {Phys. Rev.
  Lett.}\ }\textbf {\bibinfo {volume} {109}},\ \bibinfo {pages} {118305}
  (\bibinfo {year} {2012})}
\bibitem [{\citenamefont {Olmsted}(2008)}]{olmsted}%
  \bibfield  {author} {\bibinfo {author} {\bibfnamefont {P.~D.}\ \bibnamefont
  {Olmsted}},\ }\href@noop {} {\bibfield  {journal} {\bibinfo  {journal}
  {Rheologica Acta}\ }\textbf {\bibinfo {volume} {47}},\ \bibinfo {pages} {283}
  (\bibinfo {year} {2008})}
\bibitem [{\citenamefont {Da~Cruz}(2004)}]{dacruz}%
  \bibfield  {author} {\bibinfo {author} {\bibfnamefont {F.}~\bibnamefont
  {Da~Cruz}},\ }\emph {\bibinfo {title} {Ecoulements de Grains Secs: Frottement
  et Blocage}},\ \href@noop {} {Ph.D. Thesis},\ \bibinfo  {school} {Ecole des
  Ponts ParisTech} (\bibinfo {year} {2004})
\bibitem [{\citenamefont {Ball}\ and\ \citenamefont {Melrose}(1995)}]{melrose}%
  \bibfield  {author} {\bibinfo {author} {\bibfnamefont {R. C.}~\bibnamefont
  {Ball}}\ and\ \bibinfo {author} {\bibfnamefont {J. R.}~\bibnamefont {Melrose}},\
  }\href@noop {} {\bibfield  {journal} {\bibinfo  {journal} {Adv.
  Colloid and Interface Sci.}\ }\textbf {\bibinfo {volume} {59}},\ \bibinfo
  {pages} {19} (\bibinfo {year} {1995})}
\end{thebibliography}
\end{document}